\documentclass[aps,pra,reprint]{revtex4-1}

\usepackage{graphicx}
\usepackage{color}
\usepackage{amsmath}
\usepackage{amssymb}

\begin{document}
\title{Exotic domain walls in Bose-Einstein condensates with double-well dispersion}
\author{Tongtong Liu}
\author{Logan W. Clark}
\author{Cheng Chin}
\affiliation{The James Franck Institute, Enrico Fermi institute and Department of Physics, \\ The University of Chicago, Chicago, IL 60637, USA}

\date{\today}

\begin{abstract}
We study the domain walls which form when Bose condensates acquire a double-well dispersion.
Experiments have observed such domain walls in condensates driven across a $\mathbb{Z}_2$ symmetry-breaking phase transition in a shaken optical lattice. We derive a generic model to describe the dispersion and to compute the wavefunctions and energies of the domain walls. We find two distinct regimes which demand different physical pictures. In the weak coupling regime, where interactions are weak compared to the kinetic energy barrier, ``density wave domain walls'' form that support an extended density wave and a series of phase steps. These features can be understood as the quantum interference between domains with distinct momenta. In the strong coupling regime where interaction dominates, the system forms ``phase domain walls'' which have the minimum width allowed by the uncertainty principle and suppressed density modulation. Analytic results for the domain wall wavefunctions are obtained in the two regimes. The energy of domain walls behaves similarly to that of topological defects in paradigmatic field theories.
\end{abstract}


\maketitle

\bibliographystyle{apsrev4-1}

\section{Introduction}

Topological defects arise in many-body systems in fields as diverse as cosmology, particle physics, superfluidity, quantum field theory, liquid crystals and metallurgy \cite{toporeview, Vilenkin2000}; examples include cosmic strings, spin domain walls, magnetic monopoles, superfluid vortices, and dislocations. In all of these forms, topological defects  are supported by boundary conditions homotopically distinct from the vacuum, and are thus robust against any local operation. Aside from being fascinating objects in their own right, topological defects may hold the key to enigmas in observational cosmology \cite{Vilenkin2000}, the quantum origin of objects which behave macroscopically \cite{Blasone2011}, and even the presence of gapless modes in topological matter \cite{Teo2010}.

Topological defects often form in the symmetry-breaking process across a phase transition. In field theory the $\phi^4$ and sine-Gordon equations are paradigmatic models that support such defects and offer insight into their universal behavior \cite{Jackiw1977}. In condensed matter physics the quantum Ising-model \cite{Sachdev2011} describes a $\mathbb{Z}_2$ symmetry-breaking quantum phase transition which generates domain walls separating ferromagnetic domains \cite{KZM2, KZM3, KZM4}. The formation of topological defects has been widely studied in the framework of the ``Kibble-Zurek'' mechanism \cite{Kibble, zurek, delCampo2014} in systems including the early universe \cite{Kibble}, superfluid helium \cite{zurek}, liquid crystals \cite{Chuang1991}, and recently in cold atomic gases \cite{Sadler2006,Weiler2008,Baumann2011,Chen2011,Corman2014,Navon2015,Anquez2015,Donadello2016,Clark2016}.

\begin{figure}
\includegraphics[width=2.8 in]{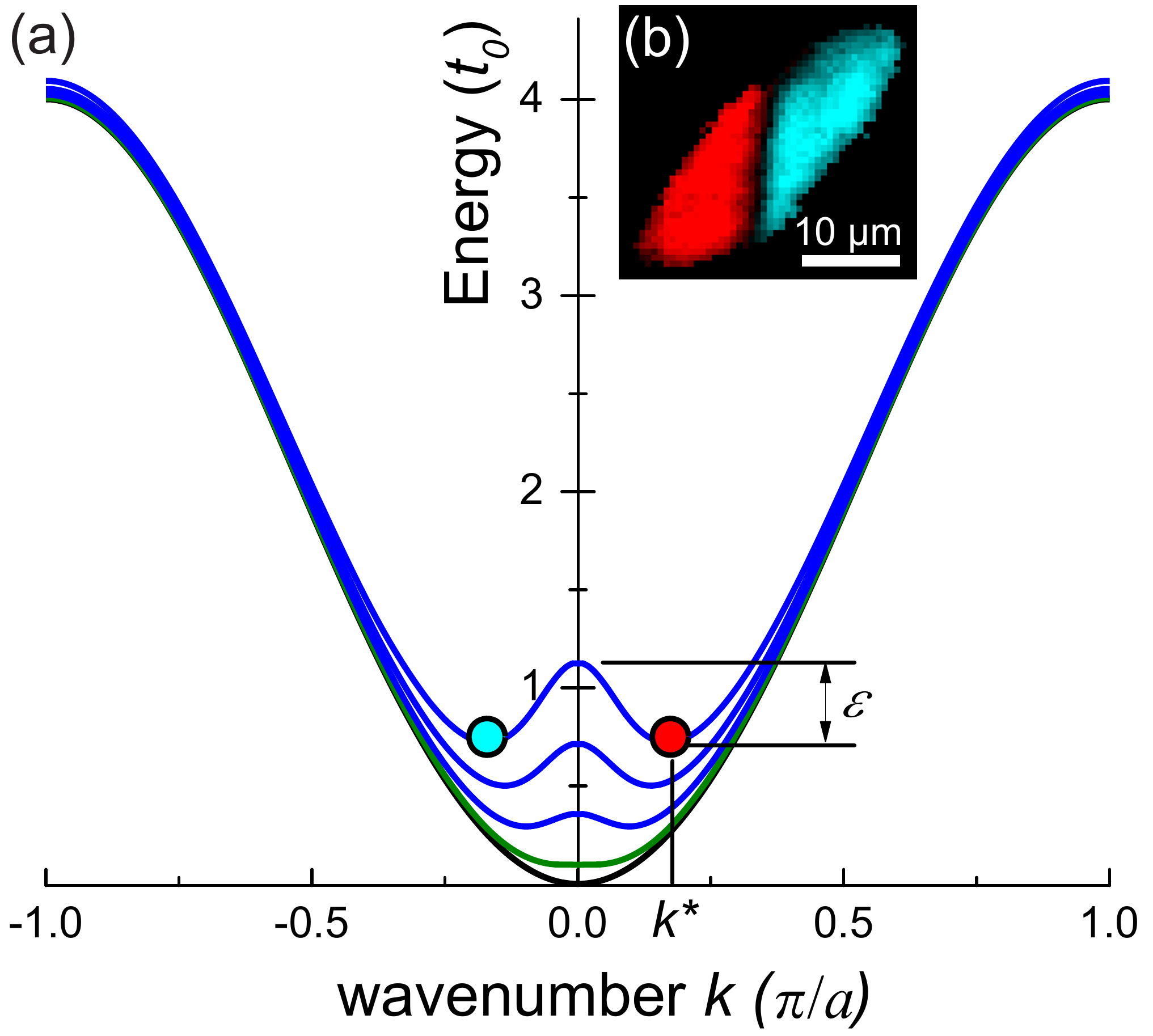}
\caption{\label{fig:1} (Color online) Emergence of double-well dispersion for atoms in a shaken optical lattice. (a) Ground band structures for atoms in a shaken lattice of spacing $a$ calculated by diagonalizing Eq.~(\ref{eqn:Hsingle}) have a single-well structure for small shaking amplitudes (black curve),
become quartic at $k=0$ at the critical shaking amplitude (green curve), and develop a double-well structure for large shaking amplitudes (blue curves). The double-well dispersion is characterized by the kinetic energy barrier $\varepsilon$ and the wavenumbers at the minima $k=\pm k^*$. (b) An averaged image of Bose condensates with a single domain wall. The color represents atoms condensed to wavenumber $k=k^*$ (red) and $k=-k^*$ (cyan).}
\end{figure}

Recent experiments revealed a novel topological defect in a Bose-Einstein condensate with double-well dispersion \cite{Parker2013, Clark2016}. The experiments show that shaking a condensate with an optical lattice induces a $\mathbb{Z}_2$ symmetry-breaking phase transition when the dispersion $E(k)$ evolves from a single-well at wavenumber $k=0$ to a double-well with two minima at $k=\pm{k^*}$, see Fig.~\ref{fig:1}(a). After the phase transition bosons condense at one of the two minima, breaking the inversion symmetry ($k\rightarrow -k$) of the Hamiltonian. The locations of domain walls formed during the phase transition follow a universal space-time scaling symmetry \cite{Clark2016}. An example image of a Bose condensate with a domain wall is shown in Fig.~\ref{fig:1}(b).

In this paper we study the domain wall wavefunction and energy in a Bose condensate with double-well dispersion. First, in Sec.~\ref{sec:theory} we derive a model for condensates with domain walls based on the Gross-Pitaevskii equation (GPE). Using that model, in Sec.~\ref{sec:wavefunctions} we present the wavefunctions over a wide range of coupling strengths and find that the domain walls develop universal structures for weak and strong coupling which demand distinct physical pictures. In Sec.~\ref{sec:energy} we compute the excitation gap of domain walls, including its contributions from kinetic energy, interactions, and mass. Finally, in Sec.~\ref{sec:conclusion} we conclude and discuss the relevance of these results to experiments.

\section{Theoretical Model} \label{sec:theory}

The development of double-well dispersion in the experiments can be best understood in terms of the hybridization between single atom Bloch bands \cite{Parker2013,zhaihui}. We consider an optical lattice of spacing $a$ which is periodically displaced at the frequency $\omega$ tuned near the band gap $G=E_1(0)-E_0(0)$ between the ground band energy $E_0(k)$ and the first excited band energy $E_1(k)$ at wavenumber $k=0$. The modulation couples these two bands, whereas the couplings to other bands are off-resonant and therefore negligible. Under the rotating-wave approximation, the hybridization of the two bands can be modelled by the single particle Hamiltonian

\begin{equation}
\label{eqn:Hsingle}
H_\mathrm{s}(k)=\left[\begin{array}{cc}
        E_1(k) & 0 \\
        0 & E_0(k) \\
      \end{array}\right]-\frac{\hbar\omega}2\sigma_z+\frac{\Omega}2\sigma_x,
\end{equation}

\noindent where $\sigma_z$ and $\sigma_x$ are the Pauli matrices, and $\Omega$ characterizes the mixing of the two bands. For a small positive detuning $\Delta=\hbar\omega-G>0$, the ground band dispersion develops a double-well structure when the mixing $\Omega$ exceeds a critical value $\Omega_\mathrm{c}$, see Fig.~\ref{fig:1}(a). The dispersion is then characterized by two minima at $k=\pm k^*$ and a kinetic energy barrier of height $\varepsilon$ at $k=0$, such that $\pi/k^*$ and $\varepsilon$ serve as natural length and kinetic energy scales, respectively.

Our calculations to determine the wavefunctions and energies of domain walls are based on the GPE, which provides a mean-field description of an atomic condensate \cite{textbook}. In a weak lattice the GPE is applicable because the condensate remains a superfluid, and the shaken lattice modifies the kinetic energy term. Near the critical point where the double-well dispersion emerges, the three-dimensional (3D) GPE for a homogeneous condensate with a one-dimensional (1D) lattice along the $x$-axis can be written as

\begin{equation}
(\alpha\partial_x^2+ \beta\partial_x^4 + \frac{\hbar^2\mathbf{\hat{k}}^2_{\bot}}{2m}+g_3|\Psi|^2)\Psi=\mu\Psi,
\label{eqn:3DGPE}
\end{equation}

\noindent where $\mathbf{\hat{k}_{\bot}}=\hat{k}_y\mathbf{e_\mathrm{y}}+\hat{k}_z\mathbf{e_\mathrm{z}}$ is the wavenumber operator perpendicular to the lattice, $m$ is the atomic mass, $g_3$ is the coupling strength in 3D, and $h\equiv2\pi\hbar$ is the Planck constant. As shown below, the parameters $\alpha$ and $\beta$ can be extracted from the hybridized ground band energy of the two-band model in Eq.~(\ref{eqn:Hsingle}) near the critical point $\Omega=\Omega_\mathrm{c}$.

Assuming that the domain wall is perpendicular to the lattice ($x$-axis) and that the wavefunction has no excitation along the $y$- or $z$-directions, we can simplify Eq.~(\ref{eqn:3DGPE}) to a 1D GPE along the lattice by substituting $\Psi(x, y, z)\equiv \psi(x)/\sqrt{A}$ to obtain

\begin{equation}
(\alpha\partial_x^2+\beta\partial_x^4+g|\psi|^2)\psi=\mu\psi,
\label{eqn:1DGPE}
\end{equation}

\noindent where $g\equiv{g_3/A}$ is the coupling strength in one dimension and $A$ is the area of the system transverse to the lattice. While experiments have focused on 3D condensates, this equation can be considered to represent systems of any dimension for which the mean-field assumption is valid and there are no excitations transverse to the lattice.

Using $E_0(k)=2t_0[1-\cos(ka)]$ and $E_1(k)=G-2t_1[1-\cos(ka)]$ in the tight binding regime, where $t_0$ and $t_1$ are the ground and first excited band tunnelling energies, respectively, we obtain in the critical regime $|\Omega-\Omega_\mathrm{c}|\ll\Omega_\mathrm{c}$:

\begin{eqnarray}
\alpha  &=& \frac{\sqrt{t_0t_1}(t_1-t_0)^2}{(t_0+t_1)^2}\frac{a^2}{\Delta}(\Omega-\Omega_\mathrm{c}),\\
\beta   &=& \frac{t_0t_1(t_1-t_0)}{t_0+t_1}\frac{a^4}{\Delta},\\
\Omega_\mathrm{c}&=& \frac{2\sqrt{t_0t_1}}{t_1-t_0}\Delta.
\end{eqnarray}

We can further derive the characteristic wavenumber $k^*$ and the kinetic energy barrier $\varepsilon$ for $\Omega>\Omega_\mathrm{c}$, which are given by

\begin{eqnarray}
k^*&=&\sqrt{\frac{\alpha}{2\beta}}=\frac1a \sqrt{\frac{\Delta}{t_1+t_0}}(\frac{\Omega-\Omega_\mathrm{c}}{\Omega_\mathrm{c}})^{1/2}, \label{eqn:kstar}  \\
\varepsilon&=&\frac{\alpha^2}{4\beta}=\Delta\frac{t_1t_0(t_1-t_0)}{(t_1+t_0)^3}(\frac{\Omega-\Omega_\mathrm{c}}{\Omega_\mathrm{c}})^2.
\end{eqnarray}

\noindent Here the exponent $1/2$ in Eq.~(\ref{eqn:kstar}) characterizes the divergence of the length scale $\pi/k^*$ near the critical point.

We numerically solve Eq.~(\ref{eqn:1DGPE}) in the dimensionless form:

\begin{equation}
(1+2\partial_u^2+\partial_u^4+|\varphi(u)|^2)\varphi(u)=\frac{\mu}{\varepsilon}\varphi(u),
\label{eq:1nD}
\end{equation}

\noindent where we have introduced the dimensionless length $u\equiv{k^*x}$ and wavefunction $\varphi(u)\equiv(g/\varepsilon)^{1/2}\psi(x)$. An offset of $1$ is added to the kinetic energy in order to conform with the convention that the chemical potential of a non-interacting condensate is zero. Here the dimensionless ratio $\mu/\epsilon$ is the only parameter and characterizes the coupling strength of the system. The ground state wavefunction is thus given by  $\varphi_0(u)=(\mu/\varepsilon)^{1/2}\exp(\pm i u)$.

We determine the stationary state wavefunction that supports a domain wall by solving Eq.~(\ref{eq:1nD}) with the imaginary time evolution method. We introduce an initial ansatz with a single domain wall at $u=0$ while satisfying the boundary conditions $\partial_u\varphi(u\rightarrow\pm\infty)=\pm{i\varphi}$ far from the domain wall. We choose a grid spacing of $\Delta u=0.01$ and a range of up to $-500\leq u\leq 500$ in order to resolve and contain the full spatial structure of the wavefunction with negligible errors. We evolve the time-dependent GPE, which corresponds to Eq.~(\ref{eq:1nD}) with $\mu$ replaced by $i\hbar\partial_t\equiv\partial_{\tau}$, using the split-step Crank-Nicolson method \cite{dautray2012} in imaginary time $\tau=-i t/\hbar$ \cite{chiofalo2000}.
Derivatives at the edge of the wavefunction are given by the boundary conditions, which enforce the presence of a single domain wall. The wavefunction is normalized after each interval to keep the asymptotic density constant. Eventually, the solution settles to a stationary wavefunction that supports a single domain wall. We have checked that our results are insensitive to the choice of the initial trial functions. For clarity, all results in the remainder of this work are presented in dimensional units.

\section{Domain wall wavefunctions} \label{sec:wavefunctions}

\begin{figure}
\includegraphics[width=2.8 in]{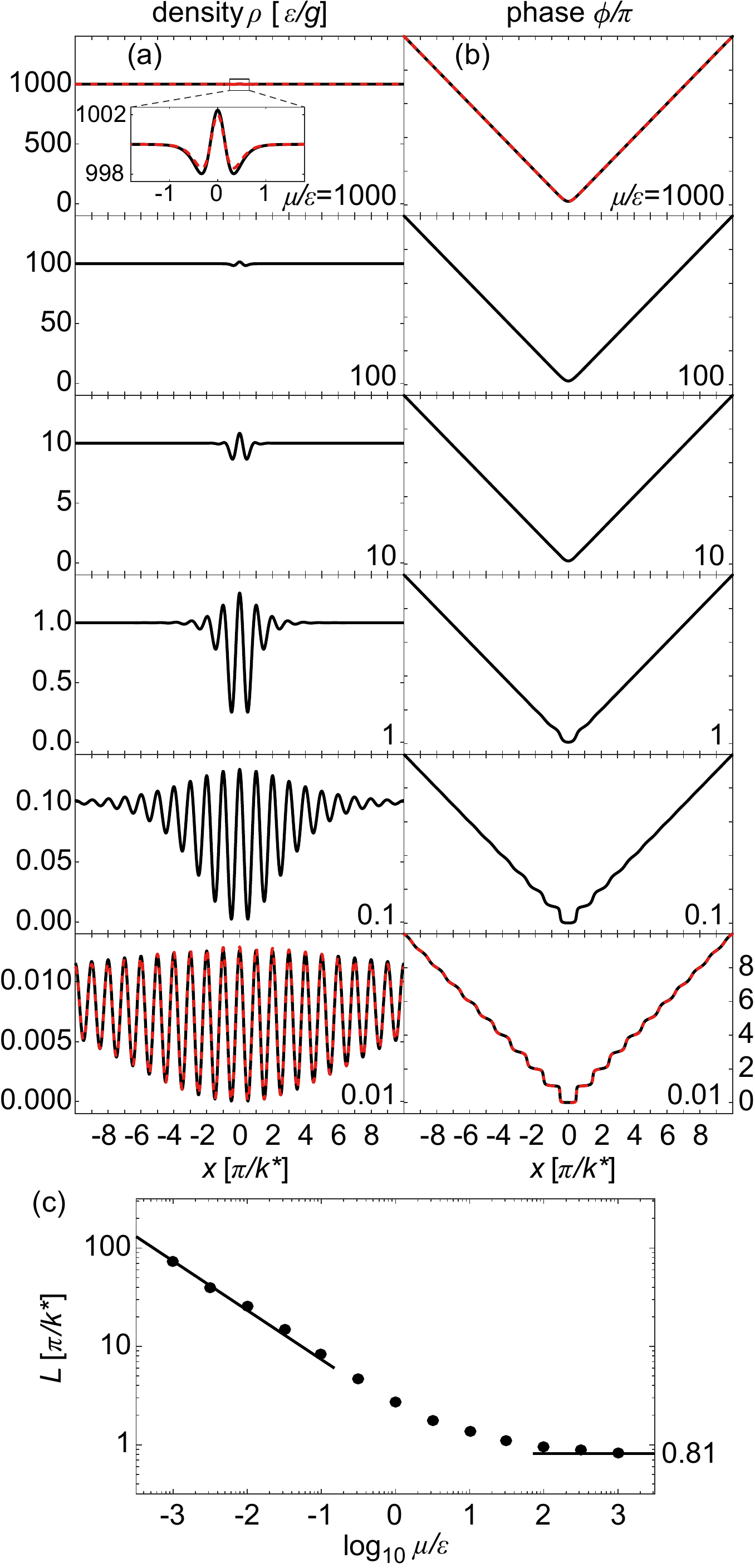}
\caption{\label{fig3} (Color online) Wavefunctions for domain walls at different coupling strengths. The (a) density $\rho$ and (b) phase $\phi$ of the condensate wavefunctions $\psi=\sqrt{\rho}e^{i\phi}$ with a domain wall at $x=0$ vary dramatically with the coupling strength $\mu/\varepsilon$. Away from the domain wall, the asymptotic density and phase are $\rho_{\infty}=\mu/g$ and $\phi=\pm k^*x$, respectively. In the weak coulping regime $\mu/\varepsilon\ll1$, the period of the density oscillations and the step size of the phase jumps are $\pi/k^*$ and $\pi$, respectively. Red dashed curves show the analytic wavefunctions given in Eq.~(\ref{psi_weak}) in the weak coupling regime and Eq.~(\ref{psi_strong}) in the strong coupling regime. (c) The width of the domain wall $L$. The solid curves indicate the asymptotic behaviors $L\propto(\mu/\epsilon)^{-1/2}/k^*$ for weak coupling and $L\approx0.81\pi/k^*$ for strong coupling.}
\end{figure}

The calculated wavefunctions across a domain wall for different coupling strengths $\mu/\epsilon$ are shown in Fig.~\ref{fig3}. We extract three key quantities from the wavefunctions: the density $\rho(x)$, the phase $\phi(x)$, and the width $L$. The width of the domain wall $L\equiv 2\pi/\Delta{k}$ is defined based on the full width at half maximum $\Delta{k}$ of the peak near $k=2k^*$ in the Fourier transform $F(k)=\hat{\mathcal{F}}\rho(x)$ of the density. 

Striking features of the domain wall wavefunctions distinguish two important regimes. In the weak coupling regime $\mu/\varepsilon\ll1$ the domain wall supports an extended density wave, whereas in the opposite strong coupling regime $\mu/\varepsilon\gg1$ the width $L$ and the density variation of the domain wall are strongly suppressed, see Fig.~\ref{fig3}. In both regimes, the density at the domain wall does not reach zero and thus the two domains are phase coherent. We note that, since the kinetic energy scale $\varepsilon$ vanishes at the critical point, the strong coupling regime also corresponds to the quantum critical regime for a condensate with a finite chemical potential. In the following, we investigate the properties of domain walls in each regime in more detail.

The weak coupling regime is distinguished by the large width over which the condensate exhibits oscillations in density. At the same time, the phase of the wavefunction supports intriguing quantized steps of $\pi$. Both the density oscillations and phase steps have a characteristic length scale of $\pi/k^*$ and become increasingly prominent as $\mu/\varepsilon$ reduces toward zero.

We use a two-component picture to capture the key features of the wavefunctions in the weak coupling regime,

\begin{equation}
\psi(x)\approx\sqrt{\rho_+(x)}e^{ik^*x}+\sqrt{\rho_-(x)}e^{-ik^*x},
\label{psi_weak}
\end{equation}

\noindent where $\rho_\pm$ are the density envelopes of the positive/negative momentum components $(k=\pm k^*)$ shown in Fig.~\ref{fig4}(c). The density envelopes can be approximated by (see Appendix~\ref{app:weak})

\begin{equation}
\rho_{\pm}(x) \approx\frac{\mu}{4g}[1 + \tanh(\pm \frac{x}{\xi} + d)]^2,
\label{weak_analytical_approx}
\end{equation}

\noindent where $d=0.134$ is a constant and $\xi=3.39(\varepsilon/\mu)^{1/2}{k^*}^{-1}$ relates to the width of the domain wall $L$. The values of $d$ and $\xi$ are chosen to minimize the error of this approximation near $x=0$. The wavefunction given by Eqs.~(\ref{psi_weak}) and (\ref{weak_analytical_approx}) reproduces the calculations for $\mu/\varepsilon\le0.01$ with an agreement of $>96\%$ over the entire domain, see bottom row of Fig.~\ref{fig3}(a) and (b). Remarkably, in this limit the density envelope is universal in the scaled coordinate $w\equiv (\mu/\varepsilon)^{1/2}k^*x$ and resembles an immiscible two-component condensate \cite{Hall1998, Stenger1998,Timmermans1998, Chui1998, Barankov2002}, see Appendix~\ref{app:weak}.

We name the domain walls in the weak coupling regime ``density wave domain walls''. The key features of the wavefunction result from interference where the two domains overlap. In particular, this explains the evenly spaced density modulation with wavenumber $2k^*$, see the lower panels of Fig.~\ref{fig3}(a). Exactly at $x=0$, constructive interference leads to a density enhancement above the background density $\rho(0) \approx 1.24 \rho_{\infty}$. Moreover, interference leads to the steps of the phase function $\phi(x)$, see the lower panels of Fig.~\ref{fig3}(b). To better visualize how interference leads to these two features, we illustrate the two-component wavefunction in the complex plane in Fig.~\ref{fig4}. The wavefunction forms a positive spiral $(\partial_x\phi=k^*)$ in one domain and a negative spiral $(\partial_x\phi=-k^*)$ in the other domain. Near the domain wall, the superposition of the spirals leads to a highly elliptic trajectory; see Fig.~\ref{fig4}(d). This trajectory corresponds to the oscillating density $\rho(x)$ as well as the quantized jumps in the phase $\phi(x)$.

The domain wall supports a density wave in the weak coupling regime due to the high kinetic energy cost of deviating from $k=\pm k^*$. The density envelopes $\rho_\pm$ must decay gently to minimize the spread of uncertainty around the minima. The competition between the residual kinetic energy from the envelope and interaction energy of the density waves determines the width of the domain wall in this regime. The interaction energy per particle at the domain wall $\mathcal{E}_\mathrm{i}$ is proportional to the chemical potential as $\mathcal{E}_\mathrm{i}\propto \mu$ while the kinetic energy per particle $\mathcal{E}_\mathrm{k}$ scales with the domain wall width $L$ as $\mathcal{E}_\mathrm{k}\propto \varepsilon(k^*L)^{-2}+\varepsilon(k^*L)^{-4}$. In the limit of $L\gg \pi/k^*$, a balance of the two energies $\mathcal{E}_\mathrm{k} \sim \mathcal{E}_\mathrm{i}$ gives $L\propto(\mu/\varepsilon)^{-1/2}/k^*$. The widths of the domain walls follow the expected scaling in the weak coupling regime, see Fig.~\ref{fig3}(c).

\begin{figure} 
\includegraphics[width=3.4 in]{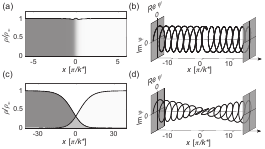}
\caption{Illustrations of the domain wall in the strong and weak coupling regimes. Strong coupling regime: (a) the calculated density profile at $\mu/\varepsilon=100$ shows a smooth transition from $k^*$ (gray) to $-k^*$ (white) across the domain wall; (b) the wavefunction plotted in the complex plane stays near the surface of a cylinder and merely switches helicity at the domain wall. Weak coupling regime: (c) the wavefunction calculated with $\mu/\varepsilon=0.01$ is well described by two domains with distinct wavenumbers $k=k^*$ (gray) and $k=-k^*$ (white) which overlap near $x=0$; (d) the wavefunction in the complex plane takes on a highly elliptical trajectory near the domain wall at $x=0$.} \label{fig4}
\end{figure}

Contrary to the weak coupling regime, in the strong coupling regime $\mu/\varepsilon\gg1$ the density $\rho(x)$ is nearly uniform and the phase function $\phi(x)$ varies smoothly across the domain wall. Since the presence of a domain wall is primarily visible only in the phase of the wavefunction, we call them ``phase domain walls'' in this regime.  The form in Eq.~(\ref{psi_weak}) is no longer adequate to describe the system in this case. Instead, we obtain an analytic solution to the GPE (Eq.~(\ref{eq:1nD})) in the strong coupling limit (see Appendix~\ref{app:strong}):

\begin{equation}
\psi(x)=\sqrt{\rho_{\infty}+\delta{\rho(x)}}e^{i\phi(x)},
\label{psi_strong}
\end{equation}

\noindent where the density variation $\delta\rho(x)$ is

\begin{equation}
\delta{\rho(x)} = \frac{\varepsilon}{g}(6 - 4\cosh2k^*x)\mathrm{sech}^4k^*x,
\label{deltan_strong}
\end{equation}

\noindent and the wavefunction phase $\phi(x)$ is

\begin{equation}
\phi(x)=\ln(e^{k^*x}+e^{-k^*x}).
\label{phi_strong}
\end{equation}

\noindent The above analytic form excellently matches the calculated wavefunctions in the strong coupling regime, see Figs.~\ref{fig3}(a) and (b) for $\mu/\varepsilon=1000$. The ``phase domain walls'' are illustrated in the complex plane in Figs.~\ref{fig4}(a) and 3(b). In this regime the wavefunction remains confined on the surface of a cylinder across the domain wall with minimal density variation.

The ``phase domain walls'' take this form because of the large interaction energy cost of density variations in this regime. Interactions are minimized by keeping the density near its background value and shrinking the domain wall. The width eventually reaches a minimum value beyond which the growth of kinetic energy with $\mu$ would overwhelm the decrease of interactions.
Indeed, the minimum width $L\approx0.81\pi/k^*$ requires the phase function $\phi(x)$ to smoothly evolve from $\partial_x\phi=-k^*$ to $+k^*$ across the domain wall due to the uncertainty principle, incurring a fixed kinetic energy cost per particle \mbox{$\mathcal{E}_\mathrm{k}\propto\varepsilon$} at the domain wall.

\section{Domain wall excitation gap} \label{sec:energy}
Beyond determining the shape of the wavefunction, the finite excitation energy gap is a key property characterizing a topological defect. For example, it plays an important role in predicting the force on defects in inhomogeneous potentials \cite{Kivshar1989, Carretero-Gonzalez2008}. Here we study the excitation energy of a domain wall quantitatively and identify its physical origins across all coupling strengths.

We define the excitation energy gap of a domain wall $E_\mathrm{g}$ based on a grand canonical ensemble as the difference of free energy $F=E-\mu N$ between systems with and without a domain wall \cite{textbook},

\begin{equation}
E_\mathrm{g}=F-F_0=\Delta E_\mathrm{k}+\Delta E_\mathrm{i}-\mu\Delta N,
\label{freeenergy}
\end{equation}

\noindent where $F$ and $F_0$ are the free energies for states with and without a domain wall, respectively. Their differences in kinetic energy $\Delta E_\mathrm{k}$, interaction energy $\Delta E_\mathrm{i}$ and particle number $\Delta N$ can be calculated from the wavefunctions with a domain wall $\psi$ and the ground state $\psi_0=\sqrt{\mu/g}e^{\pm ik^*x}$:

\begin{eqnarray}
\Delta E_\mathrm{k}&=&\int (\psi^*\hat{E_\mathrm{k}}\psi-\psi^*_0\hat{E_\mathrm{k}}\psi_0)dx,\label{eq:integralEK}\\
\Delta E_\mathrm{i}&=&\dfrac{g}{2} \int (|\psi|^4-|\psi_0|^4)dx,\\
\Delta N&=&\int(|\psi|^2-|\psi_0|^2)dx\label{eq:integralDN},
\end{eqnarray}

\noindent where the kinetic energy operator is $\hat{E}_k=\varepsilon+\alpha\partial_x^2+\beta\partial_x^4$. Note that a relationship between the energies is obtained by integrating Eq.~(\ref{eq:1nD}) for the wavefunctions $\psi$ and $\psi_0$, which yields

\begin{equation}
\Delta E_\mathrm{k}+2\Delta E_\mathrm{i}=\mu\Delta N.
\label{Erelation}
\end{equation}

\noindent Together with Eq.~(\ref{freeenergy}), we find that the excitation gap can be expressed as $E_\mathrm{g}=-\Delta E_\mathrm{i}$.

\begin{figure}
\includegraphics[width=3.3 in]{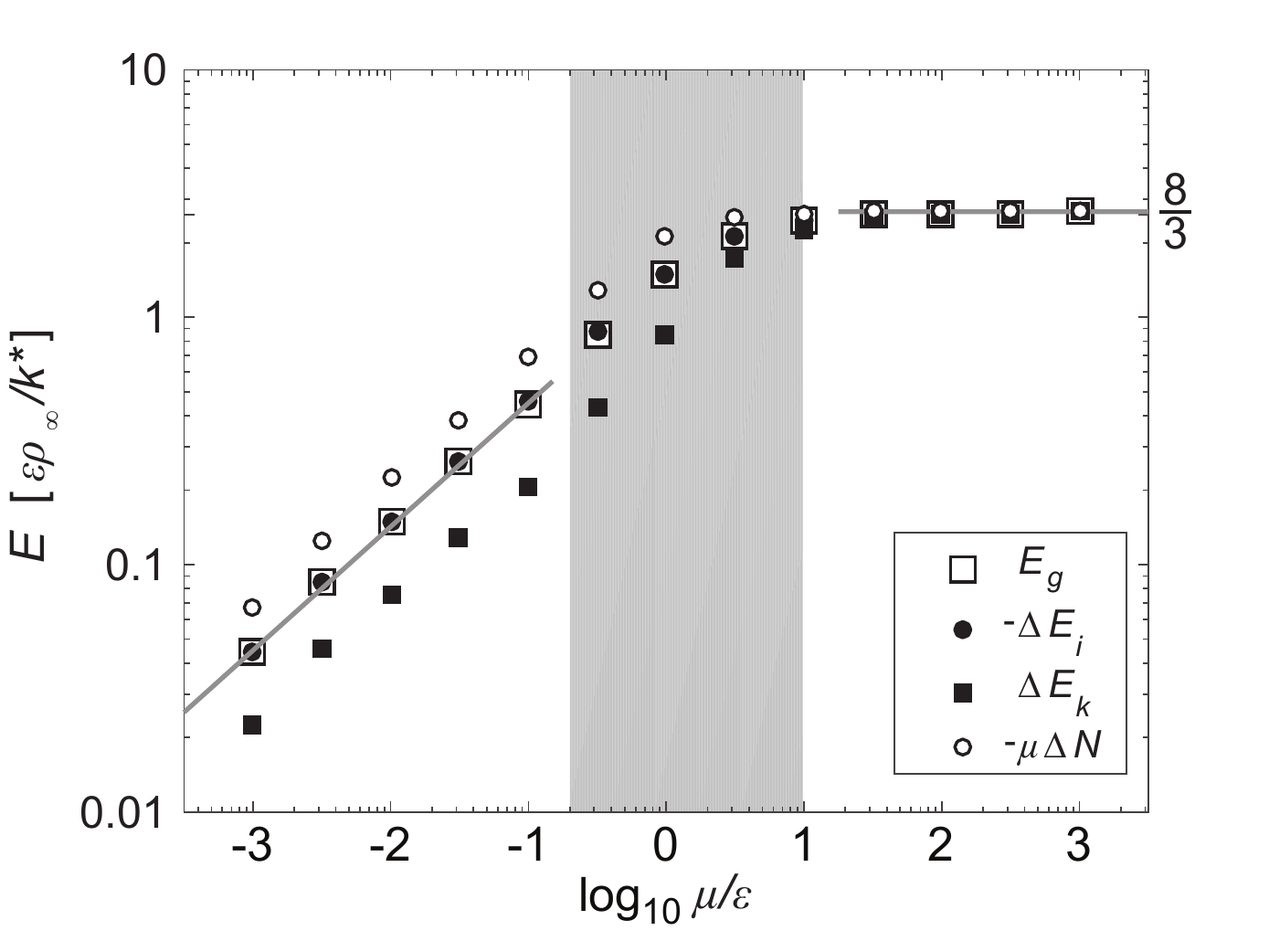}
\caption{Energy of a domain wall. Energies evaluated based on numerical calculations (symbols) are compared with asymptotic curves (gray lines) based on Eq.~(\ref{Eweak}) in the weak coupling regime and Eq.~(\ref{eqn:E_scaling_strong}) in the strong coupling regime. In dimensionless units the excitation energy $E_\mathrm{g}$ (open square), kinetic energy $\Delta E_\mathrm{k}$ (solid square), interaction energy $\Delta E_\mathrm{i}$ (solid circle) and $\mu \Delta N$ (open circles) all depend on the ratio $\mu/\varepsilon$. The gray shaded region indicates the regime that has been explored by previous experiments \cite{Parker2013, Clark2016}. Note that $\Delta E_\mathrm{i}$ and $\Delta N$ are negative.}  \label{fig5}
\end{figure}

Figure~\ref{fig5} shows the calculated energies over a wide range of coupling strengths. In the weak coupling regime our calculation shows that $E_\mathrm{g}$, $\Delta E_\mathrm{k}$, $\Delta E_\mathrm{i}$, and $\mu \Delta N$ all follow the same scaling law:

\begin{equation}
\mbox{all energies } E \propto \frac{\sqrt{\mu^{3}\varepsilon}}{gk^*}. 
\label{Eweak}
\end{equation}

\noindent A fit to the three lowest chemical potentials yields the gap $E_\mathrm{g}=c_\mathrm{w}\sqrt{\mu^3\varepsilon}/(gk^*)$ with $c_\mathrm{w}\approx1.4$. 

Intriguingly, in the strong coupling regime the energies all converge to a universal value independent of the coupling strength (Fig.~\ref{fig5}). Indeed, adopting the analytic wavefunction Eqs.~(\ref{psi_strong}-\ref{phi_strong}) we obtain from Eqs.~(\ref{freeenergy}-\ref{eq:integralDN})

\begin{equation}
\mbox{all energies } E = c_\mathrm{s}\frac{\mu\varepsilon}{gk^*}, 
\label{eqn:E_scaling_strong}
\end{equation}

\noindent where $c_\mathrm{s}=8/3$ is a constant. Excellent agreement with the numerical calculation is shown in Fig.~\ref{fig5}.

The equal magnitude of all terms in the strong coupling limit can be understood from the physical picture of the ``phase domain wall''. Given a nearly uniform density $\rho(x)$, the interaction energy is given by

\begin{equation}
\begin{split}
\Delta E_\mathrm{i} &= \frac{g}{2}\int (|\psi|^2+|\psi_0|^2)(|\psi|^2-|\psi_0|^2)dx\\
&\approx g \rho_{\infty}\int (|\psi|^2-|\psi_0|^2) dx\\
&=\mu\Delta N,
\end{split}
\label{relationinstrong}
\end{equation}

\noindent which becomes exact as $\mu/\varepsilon\rightarrow\infty$. Together with Eqs.~(\ref{freeenergy}) and (\ref{Erelation}), it can be easily shown that all energies share the same magnitude.

Even though our calculation is based on the 1D GPE, we can apply our results to a 3D condensate with a two-dimensional domain wall that is perpendicular to the shaking lattice direction. In this case, we return to the original 3D formulation in Sec.~\ref{sec:theory} and define the surface tension of a domain wall $\sigma\equiv E_\mathrm{g}/A$, which gives, from Eqs.~(\ref{Eweak}) and (\ref{eqn:E_scaling_strong}),

\begin{align}
\sigma &= c_\mathrm{w}\frac{\sqrt{\mu^3\varepsilon}}{g_3k^*} &\mbox{(Weak coupling regime)}\,\,\,\,\,\\
&= c_\mathrm{s}\frac{\mu\varepsilon}{g_3k^*} &\mbox{(Strong coupling regime)}.
\label{SurfaceTension}
\end{align}

The energy of domain walls in this system is closely related to the energy of topological defects in paradigmatic field theory models, particularly the $\phi^4$ and sine-Gordon models. In those models a classical field $\theta$ close to the vacuum expectation value satisfies a differential equation of the form (ignoring numerical constants) \cite{Jackiw1977}

\begin{equation}
-\partial_x^2\theta + \eta^2\theta^3 = M^2\theta \mathrm,
\end{equation}

\noindent where $M$ is related to the mass of the ``mesons'' and $\eta$ is the coupling constant. By comparing to Eq.~(\ref{eqn:1DGPE}) we can identify the corresponding coefficients $M\propto\sqrt{\mu}$ and $\eta\propto\sqrt{g}$. The energy of the topological defects in the field theories takes the form

\begin{equation}
E\propto M^3/\eta^2. 
\label{HEPmodels}
\end{equation} 

\noindent Remarkably, in the weak coupling regime our result Eq.~(\ref{Eweak}) for the excitation energy gap matches the form of Eq.~(\ref{HEPmodels}). Moreover, in both regimes, the domain wall energy shares the inverse dependence on the coupling constant as $E\propto\eta^{-2}$ of the field theory defects.

\section{Conclusion} \label{sec:conclusion}
The domain walls we study represent a new type of topological defect which can appear when a Bose condensate acquires a double-well dispersion. Depending on the coupling strength, we identify two kinds of domain walls: ``density wave domain walls'' in the weak coupling regime and ``phase domain walls'' in the strong coupling regime. We present two physical pictures to describe the key features of domain walls in both regimes. Both pictures provide excellent descriptions of the calculated wavefunctions and their excitation energies. We obtain analytic expressions for the wavefunctions in both regimes. 

The universal behavior of the domain wall in the strong coupling regime is particularly interesting because of its connection to quantum criticality. We find that both the wavefunction and energy associated with the domain wall take on a universal form near the critical point. Moreover, the excitation gap of a domain wall closes as $E_\mathrm{g}\sim\varepsilon/k^*\sim(\Omega-\Omega_\mathrm{c})^{3/2}$ when the system approaches the quantum critical point with constant density.

Our results are experimentally relevant as domain walls have been generated and studied both away from and close to the quantum critical regime \cite{Parker2013, Clark2016}. This work reveals exotic density structure at domain walls which should be accessible to future experiments. The length scale of the density wave is $\pi/k^*$, which is experimentally tunable from one to a few $\mu$m \cite{Clark2016} and should be visible to \textit{in situ} imaging. For ``density wave domain walls'', the contrast of the density modulation is nearly unity and the full width of the domain wall can be much larger than $\pi/k^*$. Even in experiments which cannot resolve the scale of $\pi/k^*$, the density averaged over many oscillations should still drop by about $40\%$ near the center of the domain wall.
All of these factors present favorable conditions to resolve the spatial structure of domain walls. On the contrary, for ``phase domain walls'' the observation of the density structure is more difficult because of both the narrow width and the small magnitude of the density variation. The existence of a phase domain wall is evident in the phase function $\phi(x)$ with little variation in the density profile.

This work makes an important step toward deeper investigations of these new topological objects in a quantum gas system. It enables future studies of other physical properties of domain walls, for instance the effective mass, dispersion or possible interactions between two or more domain walls. One could speculate that many domain walls supported by a Bose condensate can be modelled as quasi-particles with novel forms of dispersion and interactions. Future works could also investigate domain walls which are not perpendicular to the lattice direction, as observed experimentally in Ref.~\cite{Parker2013}, as well as in condensates with shaken lattices along multiple axes.

\section*{Acknowledgments}
We thank B. Wu and B. Anderson for helpful discussions. L.W.C. is supported by the Grainger Graduate Fellowship. This work is supported by NSF MRSEC (DMR-1420709), NSF Grant No. PHY-1511696, and ARO-MURI Grant No. W911NF-14-1-0003.

\appendix 
\section{Wavefunction in the weak coupling limit\label{app:weak}}

In the weak coupling limit $\mu/\epsilon\ll1$, we solve Eq.~(\ref{eq:1nD}) perturbatively up to the lowest order which constrains the wavefunction. Here, we expect that each domain should decay slowly across the domain wall with envelope functions $A_i$ and $B_i$ of width $L\sim(\mu/\varepsilon)^{-1/2}$ and an asymptotic density $\rho_\infty=\mu/g$. We further expect that the interference in the nonlinear term where the domains overlap is balanced by the kinetic energy from a small admixture of oscillation as $e^{\pm3iu}(\mu/\varepsilon)$ to the wavefunction. These observations motivate the ansatz

\begin{multline}
\varphi(u) = [A_1(w) e^{iu} + B_1(w) e^{-iu}]\left(\frac{\mu}{\varepsilon}\right)^{1/2} \\
+ [A_3(w) e^{3iu} + B_3(w) e^{-3iu}]\left(\frac{\mu}{\varepsilon}\right)^{3/2}, \label{eqn:weak_ansatz}
\end{multline}

\noindent where $w\equiv(\mu/\varepsilon)^{1/2}u$. 

Substituting this ansatz into Eq.~(\ref{eq:1nD}), we indeed find that terms of order $\mathcal{O}(\sqrt{\mu/\epsilon})$ cancel each other exactly without constraining the envelope functions. Therefore we keep terms up to order $\mathcal{O}((\mu/\epsilon)^{3/2})$ and obtain

\begin{align}
A_{1}&=-4\partial_{w}^{2}A_{1}+A_{1}^{3}+2A_{1}B_{1}^{2},\label{eqn:weak1}\\
B_{1}&=-4\partial_{w}^{2}B_{1}+B_{1}^{3}+2B_{1}A_{1}^{2},\label{eqn:weak2}\\
A_{3}&=\frac{1}{64}A_{1}^{2}B_{1},\label{eqn:weak3}\\
B_{3}&=\frac{1}{64}A_{1}B_{1}^{2},\label{eqn:weak4}
\end{align}

\noindent where the four equations come from terms with different wavenumbers that are independent as the width of the domain wall diverges in this limit. 

The solution of these equations represents the domain wall wavefunction to the lowest non-trivial order in $\mu/\epsilon$. Therefore, we see that in this limit the dependence of the wavefunction on the coupling strength is fully captured by the ansatz Eq.~(\ref{eqn:weak_ansatz}). 

The domain wall with weak coupling, less the interference of the two domains, is closely analogous to the boundary between different components in an immiscible two-component BEC \cite{Hall1998, Stenger1998}. Indeed, Eqs.~(\ref{eqn:weak1}) and (\ref{eqn:weak2}) are essentially identical to the GPE for two component wavefunctions $A_1$ and $B_1$ with interspecies interactions twice as strong as the intraspecies interactions \cite{Timmermans1998, Chui1998, Barankov2002}. 

Since we are unable to find a full analytical solution for Eqs.~(\ref{eqn:weak1}-\ref{eqn:weak4}), we use an analytical approximation for comparison to our numerical calculations. In particular, we drop $A_3$ and $B_3$ which make a negligible contribution to the wavefunction (though they are significant to the GPE), and use the form given in Eqs.~(\ref{psi_weak}) and (\ref{weak_analytical_approx}). There, the constants $d$ and $\xi$ are chosen to minimize error near the center of the domain wall $w\rightarrow0$. The chosen wavefunction cancels terms in the GPE up to order $\mathcal{O}(w(\mu/\epsilon)^{3/2})$.

\section{Wavefunction in the strong coupling limit\label{app:strong}}
In the strong coupling limit $\mu/\epsilon\gg1$ we can solve Eq.~(\ref{eq:1nD}) perturbatively. We start with the ansatz given in Eq.~(\ref{psi_strong}). To lowest order in the expansion parameter $\epsilon/\mu\ll1$ we find the trivial solution $\rho_\infty=\mu/g$. At this order $\phi(u)$ is unconstrained.

Continuing to the next lowest order, we obtain the differential equation

\begin{equation}
[1+2\partial_u^2+\partial_u^4+\frac{g}{\varepsilon}\delta{\rho}(u)]\varphi(u)=0.
\label{eqn:strongapp}
\end{equation}

\noindent Since derivatives of the amplitude function $\sqrt{\rho_\infty+\delta{\rho(u)}}$ are small compared to derivatives of the phase function by factors of order $\mathcal{O}(\sqrt{\mu/\epsilon})$, Eq.~(\ref{eqn:strongapp}) is dominated by the phase derivatives. Dropping the higher order terms, we obtain:

\begin{align}
0 &= \partial_u^4\phi -6\partial_u^2\phi(\partial_u\phi)^2 + 2\partial_u^2\phi, \label{eqn:app_strong_1}\\
\frac{g}{\varepsilon}\delta{\rho(u)} &= -4\partial_u^3\phi\partial_u\phi-3(\partial_u^2\phi)^2+ [(\partial_u\phi)^2 -1]^2. \label{eqn:app_strong_2}
\end{align}

\noindent 
These equations have the solution given in  Eqs.~(\ref{deltan_strong}) and (\ref{phi_strong}),
which satisfies our boundary conditions and corresponds to a condensate with a single domain wall. Intriguingly, the phase derivative $\partial_u\phi=\mathrm{tanh}(u)$ has the form of a dark soliton in an ordinary condensate.

\bibliography{ref}

\end{document}